\documentclass{aa}
\usepackage{epsfig}
\usepackage{graphicx}

\def\cm3{cm$^{-3}$}
\def\kms{km~s$^{-1}$}

\def\beq{\begin{equation}}
\def\eeq{\end{equation}}
\def\Rstar{R_{\ast}}
\def\Rsun{R_{\odot}}

\begin{document}

\title{Inferring hot-star-wind acceleration from Line Profile Variability}

\subtitle{}

\author{Luc Dessart\inst{1}
        \and
        S.P. Owocki\inst{2}
        }
\offprints{L. Dessart}

  \institute{Max-Planck-Institut f\"{u}r Astrophysik,
             Karl-Schwarzschild-Str 1, 85748, Garching bei M\"{u}nchen, Germany \\
             \email{luc@mpa-garching.mpg.de}
        \and
           Bartol Research Institute of the University of Delaware,
        Newark, DE 19716, USA \\
           \email{owocki@bartol.udel.edu}
            }

\date{Accepted/Received}

\abstract{
The migration of profile sub-peaks identified in time-monitored optical emission
lines of Wolf-Rayet (WR) star spectra provides a direct diagnostic of the
dynamics of their stellar winds via a measured $\Delta v_{LOS}/\Delta t$, a
line-of-sight velocity change per unit time. Inferring the associated
wind acceleration scale from such an apparent acceleration
then relies on the adopted intrinsic velocity of the wind material
at the origin of this variable pattern. Such a characterization of
the Line Emission Region (LER) is in principle subject to inaccuracies arising
from line optical depth effects and turbulence broadening.
In this paper, we develop tools to quantify such effects
and then apply these to reanalyze the LER properties of time-monitored
WR stars.
We find that most program lines can be fitted well with a pure optically thin
formation mechanism, that the observed line-broadening is dominated by the finite
velocity extent of the LER, and that the level of turbulence inferred through
Line Profile Variability ({\it lpv}) has only a minor broadening effect in the 
overall profile.
Our new estimates of LER velocity centroids are systematically
shifted outwards closer to terminal velocity compared to previous determinations,
now suggesting WR-wind acceleration length scales $\beta R_{\ast}$ of
the order of $10-20 \, R_{\odot}$, a factor of a few smaller than
previously inferred.
Based on radiation-hydrodynamics simulations of the line-driven-instability
mechanism, we compute synthetic {\it lpv} for C{\sc iii}5696\AA\, for WR\,111.
The results match well the measured observed migration of 20-30 m\,s$^{-2}$,
equivalent to  $\beta R_{\ast} \sim 20 \, R_{\odot}$.
However, our model stellar radius of $19 \, R_{\odot}$, typical of an O-type
supergiant, is a factor 2--10 larger than generally expected for WR core radii.
Such small radii leave inferred acceleration scales to be
more extended than expected from dynamical models of line driving,
which typically match a ``beta'' velocity law $v(r)=v_{\infty}
(1-\Rstar/r)^{\beta}$, with $\beta \approx 1-2$;
but the severity of the discrepancy is substantially reduced compared
to previous analyses.
We conclude with a discussion of how using
lines formed deeper in the wind would provide a stronger constraint
on the key wind dynamics in the peak acceleration region, while also
potentially providing a diagnostic on the radial variation of
wind clumping, an issue that remains crucial for reliable determination of
O-star mass loss rates.
}
\titlerunning{Inferring hot star wind acceleration from {\it lpv}}
\maketitle

\section{Introduction}

The hot, massive, luminous stars of spectral type O, B, and WR are
understood to have strong stellar winds driven by the line-scattering
of the star's continuum radiation
(see, e.g., Puls and Kudritzki 2000).
For OB stars, dynamical models based on the line-driving formalism of
Castor, Abbott, \& Klein (1975; hereafter CAK) and its  modern
extensions (Pauldrach, Puls, \& Kudritzki 1986; Friend \& Abbott 1986)
yield predictions for the mass loss rate ${\dot M}$ and terminal speed
$v_{\infty}$ that are generally in quite good agreement with
observationally inferred values (Puls et al. 1996; Crowther et al. 2002).
Indeed, for these relatively optically thin winds, the formation of line
profiles throughout the flow acceleration region yields a constraint on
the full wind velocity law, commonly characterized through the
phenomenological ``beta-law'' form
$v(r) = v_{\infty} (1-\Rstar/r)^{\beta}$,
where $\Rstar$ is the (hydrostatic) surface value of the radius $r$,
and the velocity exponent $\beta$ is inferred from line profile fits.
Careful time-monitoring of optical emission lines
(e.g. He II 4686; Eversberg, L\'epine, \& Moffat 1998)
show that these winds are not strictly steady, but likely consist of
quite extensive small-scale structure.
This leads to an intrinsic, low-level, Line Profile Variability ({\it lpv}),
characterized by moving subpeaks that provide an even more direct
diagnostic of the wind acceleration.
For OB stars, both the single exposure line profiles and such {\it lpv}
subpeak monitoring suggest a velocity-law exponent $\beta \approx
1-1.2$,
which is roughly consistent with the value $\beta \approx 0.8-1$
predicted from dynamical models of line-driving.


Over evolutionary timescales, the cumulative depletion of the
hydrogen-rich stellar envelope of such OB stars is understood to lead
to the Wolf-Rayet (WR) stage, characterized by reduced or no hydrogen,
with spectra dominated by broad emission lines of He and CNO processed material.
For such WR stars,
the very large mass loss rates inferred from the
strength of these wind-broadened emission lines are difficult to
reconcile with a detailed, dynamical model based on the paradigm of
radiative driving via line-scattering.
Because the higher density makes the inner wind optically thick even in the
electron scattering continuum, there are now no direct diagnostics of
the initial acceleration from the hydrostatic stellar core, making
it difficult to constrain dynamical models.
Moreover, the overall shape of the emission profile turns out to be
relatively insensitive to even the acceleration scale in the outer
part of the wind (see Sects. 3.1--3.2).

For such WR winds, the monitoring of moving subpeaks from {\it lpv} thus provides a
crucial diagnostic of the outer wind acceleration (Robert 1992).
The most extensive previous analysis of such {\it lpv} is that of L\'epine \&
Moffat (1999; hereafter LM).
By combining the measured subpeak acceleration with an estimate of
the characteristic velocity and range of the Line Emission Region
(LER), they inferred wind acceleration scales
$\beta \Rstar \approx 20  - 80 \, \Rsun$.
If applied to typical OB supergiant radii
$\Rstar \approx 20  - 45 \, \Rsun$
(Crowther et al. 2002; Hillier et al. 2003),
such acceleration scales imply a velocity index ($\beta \approx
1-2$), which is nearly compatible with the value predicted by wind models.
But both evolutionary (Langer et al. 1994; Schaerer \& Maeder 1992) and
spectroscopic (Hillier \& Miller 1999) models of WR stars suggest a
much smaller hydrostatic core radius,
$\Rstar \approx 2 - 5 \, \Rsun$,
for which the LM results imply a very large ratio between
acceleration scale to core radius, formally $\beta \approx 4 - 40$, with a typical
value $\beta \approx 10$.
Over an associated factor 10 range in  radius the radiative flux declines
by a factor of 1/100, making it difficult to understand how such extended
acceleration scales could be maintained by a radiative driving mechanism.

This difficulty motivates our present reexamination of the basic
inference of wind acceleration scales from {\it lpv}, with particular
emphasis on the assumed velocity and width of the LER.
We first provide a simple overview (Sect. 2) of how a
measured acceleration can be related to an characteristic acceleration
scale through an associated velocity range.
We next (Sect. 3) review the sensitivity of the overall shape of
emission profiles to the wind and LER parameters, showing thereby that
the observed emission line broadenings are consistent with relatively
large and extended velocities for the LER, without need to invoke large
turbulent speeds.
We then (Sect. 4) show that such larger and more extended LER velocities imply
substantially lower acceleration scales ($\beta \Rstar \approx 10-20 \, R_{\odot}$), which
are now much closer to being compatible with dynamical models that
account for the multi-line scattering expected in dense, WR winds.
In Sect. 5 we examine how well {\it lpv} and associated subpeak accelerations can
be reproduced through dynamical simulations of the intrinsic
instability of line-driving.
Finally, Sect. 6 summarizes our results and provides an outlook for
future work.

\begin{figure}
\vspace{7cm}
\includegraphics{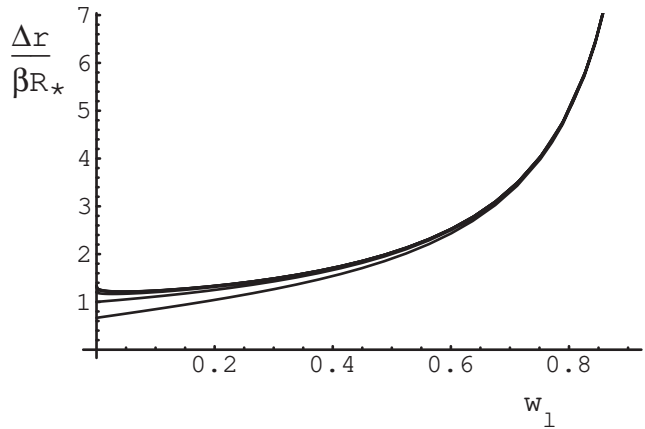}
\caption{The radial distance to reach from a given scaled velocity
$w_{l}$ to half-way to the terminal speed, $(1+w_{l})/2$, scaled by
$\beta \Rstar$, and plotted vs. $w_{l}$ for various velocity indices
$\beta = 1/2$ (lowermost curve) and $\beta=$1, 2, 4, 6, 8, and 10
(overplotted curves).
}
\end{figure}

\section{Dynamics of wind acceleration}

\subsection{Acceleration length scale}
%

To gain a basic understanding of the factors influencing an inferred
wind acceleration scale, consider first the simple case of constant
acceleration $a$ between a starting velocity $v_{l}$ and final velocity
$v_{\infty}$.
By simple dynamics, the associated length is
\beq
s_{l} = { v_{\infty}^{2} - v_{l}^{2} \over 2 a } \, .
\eeq
For application to a stellar wind, let us scale the starting velocity
by the terminal value, $w_{l} \equiv v_{l}/v_{\infty}$, and the length
by a solar radius,
\beq
{s_{l} \over \Rsun} = {1-w_{l}^{2} \over 2 {\tilde a}}
\label{slsun}
\, ,
\eeq
where ${\tilde a} \equiv a \Rsun/v_{\infty}^{2}$ represents a
convenient scaling for the acceleration.
Applying now the  acceleration magnitudes inferred by LM from WR {\it lpv},
${\tilde a} \approx 0.001 - 0.005 $, we immediately derive
associated length scales
\beq
{s_{l} \over \Rsun} = 100 - 500 \, \left ( 1-w_{l}^{2} \right )
\, .
\eeq
If we now associate the quantity $w_{l}$ defined above
with the characteristic fractional velocity of the inner edge of the
LER, we see that the typical LER speeds assumed
by LM, $w_{l} \approx 0.6$ imply very large acceleration scales,
$s_{l}/\Rsun \approx 60 - 300$.
By contrast, if we instead take $w_{l} \approx 0.9$
(see Sect. 3.2, Fig. 4), we find $s_{l}/\Rsun \approx 20 - 100$.

An alternative way to characterize the overall wind acceleration is to make
use of the phenomenological beta-velocity law
\beq
w(r) \equiv v(r)/v_\infty = (1 - \Rstar/r)^\beta
\, ,
\label{betalaw}
\eeq
for which the scaled acceleration vs. velocity  becomes
\beq
{\tilde a} = R_\odot w w' = {\beta R_\odot \over R_\ast} (1- w^{1/\beta})^2 w^{2-1/\beta} .
\eeq
However, since $w<1$, note that for moderately large $\beta$,
\beq
w^{1/\beta} = e^{(1/\beta) \ln w} \approx 1 + (1/\beta) \ln w \, ,
\eeq
which implies
\beq
{\tilde a}
\approx { R_\odot \over\beta \Rstar} \, ( w \ln w)^2 \, .
\eeq
This very useful property, which was first derived by  L\'epine (1998),
implies that a plot of acceleration vs. speed just scales homologously
with the parameter $\beta \Rstar$ (see Fig. 8 below).
For an assumed scaled velocity of the LER, $w_{l}$, this now gives a
scaled acceleration length,
\beq
{\beta \Rstar \over  R_\odot} \approx  { ( w_{l} \ln w_{l})^2 \over {\tilde a} }
\, .
\eeq
Comparison with Eqn. (\ref{slsun}) show that the beta-law
characterization of the acceleration scale differs from that for
constant acceleration by the replacement of numerator $(1-w_{l}^{2})$
with $(w_{l} \ln w_{l})^{2}$.
Applying now the LM values for the velocity of the LER, $w_{l}
\approx 0.6$, we find $\beta \Rstar/\Rsun \approx 20-90$.
By contrast, taking $w_{l} \approx 0.9$ gives
$\beta \Rstar/\Rsun \approx 2-9$.

Note that these are much smaller than the simple, local scale given by
Eqn. (\ref{slsun}).
The reason for the difference is illustrated in Fig. 1, which plots
the distance to reach to half the terminal speed as a function of
initial speed, scaled by the characteristic scale $\beta \Rstar$,
and overplotted for various values of the velocity index $\beta$.
The order unity values for small initial velocities shows that
$\beta \Rstar$ is indeed a characteristic acceleration scale for the
inner regions of a beta velocity law;
but at large velocities, the local acceleration scales are actually much
larger.

\subsection{Simple overview of radiative driving}

For a steady-state, spherical wind with velocity $v(r)$ as function of
radius $r$, the net flow acceleration $a=v(dv/dr)$ results from the outward
driving against the inward pull of gravity,
\begin{equation}
v {dv \over dr} = g_{rad} - { GM \over r^{2} } \, ,
\end{equation}
where $G$ and $M$ are the gravitation constant and stellar mass.
For WR and other very luminous stars, the general belief is that
the outward driving is from stellar radiation
interacting with both line and continuum opacity of wind material.
For a star of luminosity $L$, the radiative flux is $F=L/4\pi r^{2}$,
implying that overall the radiative acceleration tends to have the
same inverse-radius-squared decline as gravity.
Since gravity moreover sets the overall acceleration scale,
it is convenient to
define the gravitationally scaled accelerations
$u' \equiv r^{2}v(dv/dr)/GM$ and $\Gamma_{rad} \equiv r^{2} g_{rad}/GM$.

For driving associated with continuum opacity $\kappa_{c}$,
$g_{rad} = \kappa_{c} F/c$,  and so
\begin{equation}
\Gamma_{c} = { \kappa_{c} L \over 4 \pi GM c} \, .
\end{equation}
In the common case that the continuum is dominated by electron
scattering opacity $\kappa_{e}$
($\approx$ 0.34 cm$^{2}$/g),
this is a constant that just reduces the effective gravity by a factor
$1 - \Gamma $, with $\Gamma$ the classical Eddington parameter.

By contrast, for line opacity, the Doppler-shift deshadowing of optically
thick lines gives line driving a dependence on the flow
acceleration, parameterized within the CAK formalism as
\begin{equation}
\Gamma_{l} = f C u'^{\alpha} \, ,
\end{equation}
where $0 < \alpha < 1$ is the CAK power-index,
$f$ is a complex function of velocity, radius, and acceleration
that accounts for the finite size of the stellar disk (see CAK Eqn.
(50)),
and $C$ is a constant set by
line-opacity, luminosity, and mass loss rate.

With these scalings, the equation of motion (9) can thus be written as
\begin{equation}
u' = fC u'^{\alpha} - (1-\Gamma) \, .
\end{equation}
For the idealized case of a point-source star, $f=1$, this equation
has no explicit radial dependence, implying that each term is just a fixed
constant.
A trivial radial integration of the scaled acceleration $u'$ from a
static lower boundary stellar radius $R_{\ast}$ thus yields the velocity law
\begin{equation}
v(r) = v_\infty (1- R_{\ast}/r)^{1/2} \, ,
\end{equation}
where the terminal speed $v_{\infty} = v_{esc} \sqrt{u'} $ is
proportional to the  effective surface escape speed,
$v_{esc} \equiv \sqrt {2 GM(1-\Gamma)/R_{\ast}}$.
The critical CAK solution with the maximal mass loss rate further
requires
\begin{equation}
1 = {\partial \Gamma_{l} \over \partial u' } \, ,
\end{equation}
which when combined with Eqn. (12) implies a critical acceleration
$u'_{c}= C_{c} = \alpha/(1-\alpha)$,
yielding $v_{\infty} =  v_{esc} \sqrt{\alpha/(1-\alpha)}$.

For more complete models that take account of the finite stellar disk
(Friend \& Abbott 1986; Pauldrach et al. 1986),
the reduction of the line-force near the star (where $f<1$) lowers both
the initial acceleration and the mass loss rate.
But as $f$ increases outward -- with $f \rightarrow 1$ at large radii,
where  the stellar radiation indeed approaches the radial streaming from a
point-source -- the lower mass loss rate allows a much stronger
acceleration ($u' \approx 6-9$!), thus yielding a faster wind
terminal speed, $v_{\infty}/v_{esc} = 2.5-3$.
The resulting velocity law is no longer analytic, but can still be well fit by
a general `beta-law' form (\ref{betalaw}),
with now $\beta \approx 0.8-1$
(Pauldrach et al. 1986).

For WR stars, the importance of line overlap severely complicates the
derivation of self-consistent dynamical solution, but it is possible
to extend the basic CAK formalism if one makes the simplifying
assumption that the lines are uniformly distributed throughout the
stellar spectrum (Friend \& Castor 1983).
Using this approach, multi-line scattering models based on Monte
Carlo simulation (Springmann 1994)
and on non-isotropic diffusion analysis
(Gayley, Owocki, \& Cranmer 1995)
both show that effective trapping of radiation in the
expanding envelope can lead to a more extended acceleration,
with an effective velocity index up to $\beta \approx 2$ in the outer
wind.

An overall point from such dynamical models based on line-driving is
that it is quite difficult to obtain a velocity law with large velocity index,
i.e. $\beta > 2$.
This is in part because line-driving itself depends on the velocity
gradient, and so becomes inefficient in the extended, slow
accelerations of a large $\beta$.
Another fundamental factor regards the inverse-radius-squared
fall-off of the radiative flux, $F \sim 1/r^{2}$.
Once the mass loss rate is tuned to allow the net outward force to be
strong enough to overcome gravity near the stellar surface, the
similar radial scaling of the radial flux and gravity tends then to
make the net outward acceleration scale with the local inward
acceleration of gravity.
As shown above for the point-star CAK model, this leads to a velocity
law with $\beta =1/2$, and the terminal speed scaling with the
surface escape speed.
Corrections due to finite-disk factor or multi-line scattering can
extend the acceleration somewhat, but only up to velocity indices of
order $\beta \approx 1-2$.
Overall then, the very large velocity indices or associated
acceleration scales inferred from apparent acceleration of subpeaks in
WR {\it lpv} are quite difficult to explain from dynamical models;
this provides a key motivation for the present reexamination of these
inferred extended acceleration scales.

\section{Line formation}

\begin{figure*}
\vspace{11cm}
\includegraphics{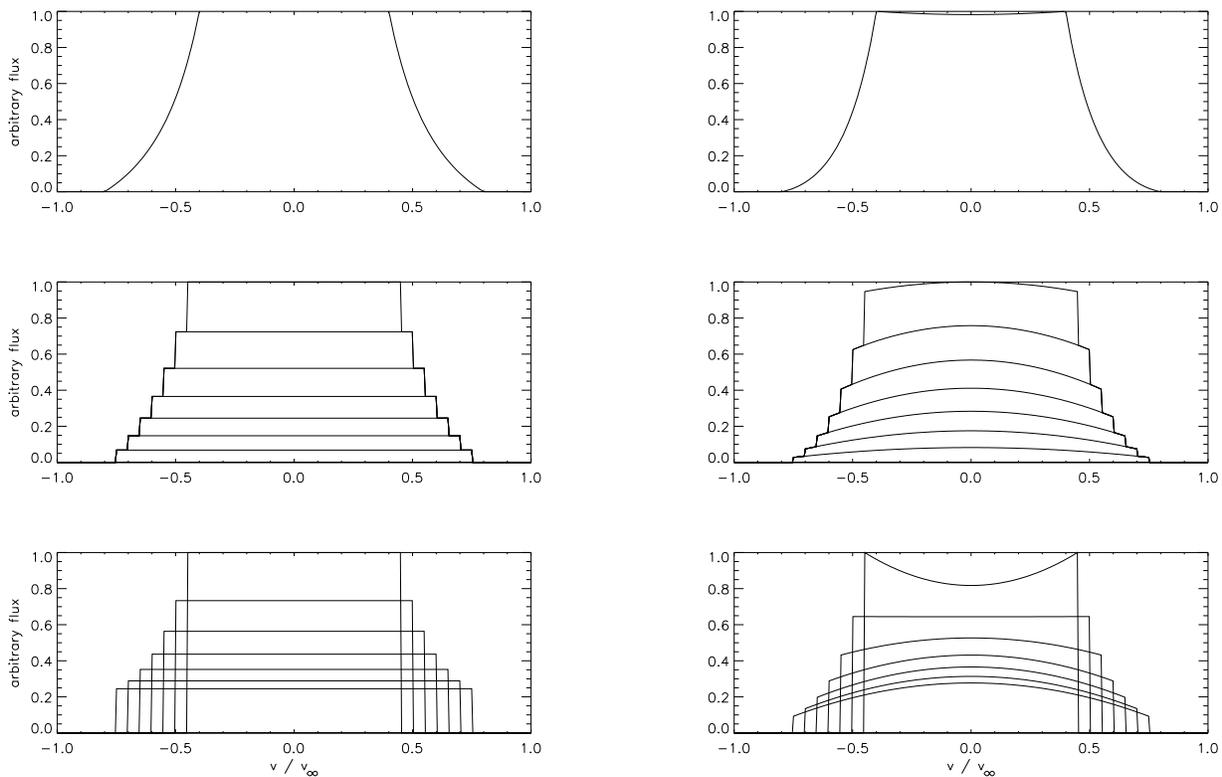}
\caption{ For optically-thin (left) and optically-thick (right) lines,
the bottom panels show the emission contribution from
seven discrete subshells with equal velocity width spanning an LER between
40\% and 80\% of $v_{\infty}$ (with $\beta=1$).
The middle panels then show a stacking of these subshell emissions to build
a cumulative line profile that approximates the spatially
integrated emission profile from the full LER, as given in the top
panel.
}
\end{figure*}

\subsection{Discrete LER without broadening or stellar occultation}

To provide a basis for the analysis of {\it lpv}, and its use as a diagnostic
of flow acceleration, let us first examine how a mean, time-steady,
line-profile depends on various wind properties.
We again assume a  spherically symmetric, monotonically
expanding wind described by the standard $\beta$ velocity law,
expressed in dimensionless form
$w(r) = v(r) / v_{\infty} = (1 - R_{\ast}/r)^{\beta}$ (Eqn. 4).
As a simple first case, in this subsection we ignore the complications of stellar occultation
and line broadening (i.e. by thermal or turbulent motions), and take the
LER to have sharp boundaries defined by the lower and upper velocities
v$_l$ and v$_u$.
The subsequent subsections will examine the effects of modifying these
simplifications.

Following Owocki \& Cohen (2001, Eqn. 7), and assuming optically-thin
emission\footnote{Here and throughout our analysis, we ignore continuum
absorption.},
the total emitted luminosity at scaled wavelength $x = (\lambda/\lambda_0 -1) c/v_{\infty} $
is given by,\beq L_{thin}(x) = 8 \pi^2 C \int_{-1}^{1} d\mu \int_{r_l}^{r_u} dr r^2 \rho^2(r)
\delta(x+\mu w(r)) \, ,
\eeq
where
$C$ is a
constant that depends on the atomic process of emission
(i.e. presently recombination).
Performing the integral over $\mu$ gives
\beq
L_{thin}(x) \propto \int_{r_l}^{r_u} dr  H[w(r)-|x|\,]  \frac{r^2 \rho^2(r)}{w(r)}
\label{lthin2}
\, ,
\eeq
where H is the Heaviside function (unity for positive argument, zero
otherwise).
Note that from a shell at radius $r$, the contribution to the profile falls
within the wavelength limits $\pm w(r)$.

Eqn. (\ref{lthin2}) applies in the case when both the line and
continuum are optically thin.
For the case of an optically thick line, the relative escape of photons along
direction cosine $\mu$ is weighted by
\beq g(\mu,r) = \frac{1+\sigma(r) \mu^2}{1+\sigma(r)/3} \, , \eeq
where
$\sigma(r) \equiv d\ln v/d\ln r -1 $.
Integration over angle then gives
(through the Dirac delta function) a corresponding weighting to the
local emissivity, yielding for the optically thick emitted luminosity
\beq  L_{thick}(x) \propto \int_{r_l}^{r_u} dr H [w(r)-|x|\,] \frac{r^2 \rho^2(r)}{w(r)}
g(|x|/w(r),r) \, .
\label{lthick}
\eeq

In the optically thin case (\ref{lthin2}), the wavelength $x$ enters only
via the Heaviside function;
thus
the emission
profile from each localized, narrow shell
(extending from $r$ to $r+dr$)
is constant within the wavelength bounds
$|x| < w(r)$ defined by the scaled velocity $w(r)$ of that shell.
By contrast, in the optically thick case (\ref{lthick}), there is an additional
wavelength dependence via the relative escape function $g$;
this gives a curvature to the shell emission profile, concave up in
the inner wind where $\sigma (r)> 0$, and concave down in the outer wind
where $\sigma (r) < 0$.

These properties of line formation are illustrated in Fig. 2.
The bottom panels show the individual flux contributions when the LER is divided
into seven discrete, finite shells spanning equal velocity bins.
In the optically thin case (left column),
the contributions are frequency independent, yielding flat-topped,
box functions.
Progressing from higher to lower velocity shells, these contributions narrow
while their emission level increases, following the increased density
in the slower, inner wind regions.
The middle panels show a cumulative stack of these individual shell contributions,
thus approximating the total emission profile of the upper panels,
which are obtained by a full radial integration over the continuum of
differential shells that span the LER.
Note that the inner velocity of the LER sets the width of the flat
top, while the outer velocity sets the maximum width of the profile
down to the continuum level.

For an optically thick line (right column), the formation follows
analogously, except that now the individual shell emission profiles
can exhibit a net curvature within the box limits.
In this particular example, wherein the LER spans the region where
$\sigma (r)$ goes from positive to negative, the individual shell
contributions go from concave up to concave down;
the net cancellation of the two opposite senses of concavity gives
the final profile a nearly flat-top form that is remarkably similar to the
optically thin case.
More generally, for optically thick lines formed predominantly in
either the inner ($\sigma > 0$) or outer ($\sigma < 0$) regions,
the total resulting profiles can exhibit either a double-peaked
(concave up) form, or a rounded (concave down) form.

In the limiting case of a constant-speed outflow, we have
$\sigma=-1$ and $g(\mu,r)=3(1-\mu^{2})/2$, yielding then upon angle
integration a scaling $1-x^{2}$ for the $g$-factor in Eqn. (\ref{lthick}),
which thus results in a parabolic shaped emission profile
(see Ignace \& Gayley 2002).

Figure 3 compares optically thin line profiles computed for
models with the same velocity range for the LER, but with much
different velocity exponents, namely $\beta =  1$ and $\beta = 4$.
Although these two values represent markedly different spatial
variations of the velocity, with much slower overall acceleration for
the $\beta = 4$ case, the resulting line profiles are remarkably
similar.
This illustrates that the mean profile shape is quite insensitive to
the wind acceleration, implying that inversion-based methods for
inferring the wind velocity law from the profile shape
(Brown et al. 1997) are nearly degenerate with respect to the
length scale of the velocity gain.

\begin{figure}
\vspace{8cm}
\includegraphics{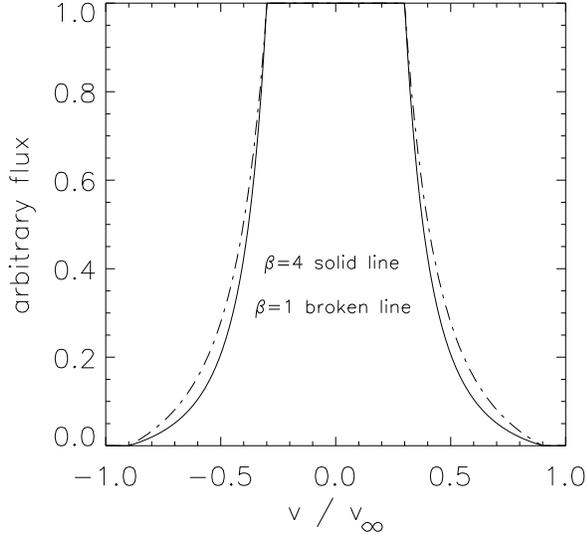}
\caption{
Synthetic emission profiles for an optically-thin line forming between
30\% and 90\% of the terminal velocity but for two different velocity laws: the
broken (solid) line corresponds to a steep (shallow) acceleration with $\beta=1$
($\beta=4$). Apart from a modified maximum line-profile flux
(not known {\it a-priori}), the dependence of the line-wing shape on $\beta$
is weak, i.e.  ca. 10\% flux-difference for an extended LER (shown here) and decreasing
for narrower LER. Inversion techniques (Brown et al. 1997) yield a very uncertain
information on the wind velocity law.
}
\end{figure}

\subsection{Gaussian LER including stellar occultation}

\begin{figure}
\vspace{8cm}
\includegraphics{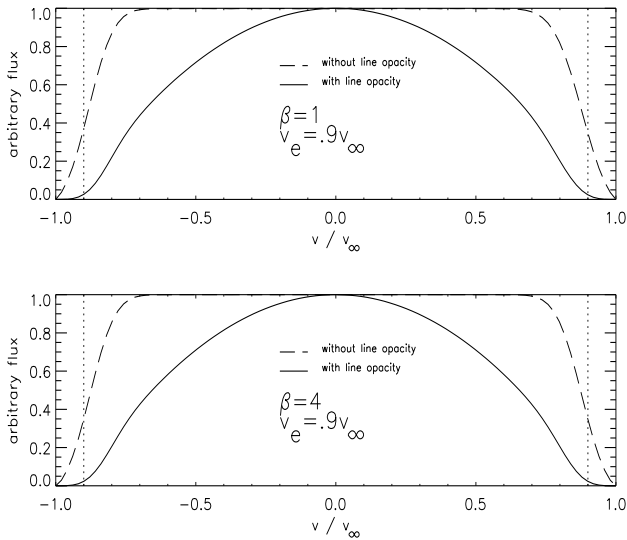}
\caption{
Study of the impact of line optical depth effects
and wind velocity law ($\beta=$  1, 4) for a synthetic line whose LER is
characterized by $v_e = .9 v_{\infty}$ and $\Delta v_e = .1 v_{\infty}$. Notice the
rounding of the line-profile which shifts its HWHM velocity $.3 v_{\infty}$ below
the adopted $v_e$. Note also the negligible impact of $\beta$ on the line-profile shape.
}
\end{figure}

\begin{figure}
\vspace{8cm}
\includegraphics{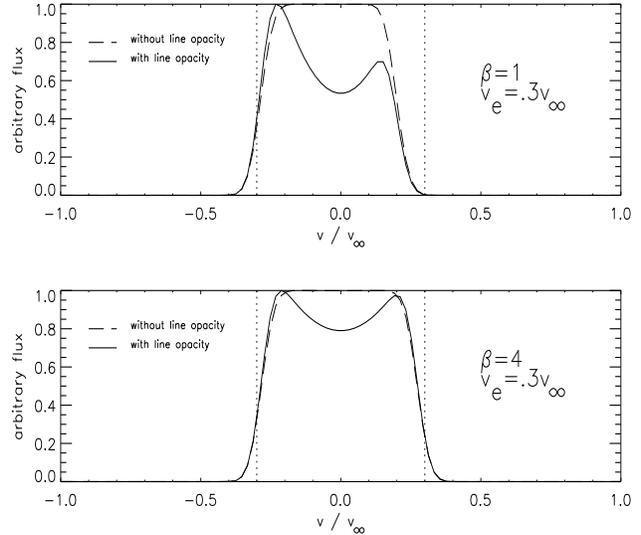}
\caption{
Same as Fig. 4, but for an inner wind line with
$v_e = .3 v_{\infty}$ and $\Delta v_e = .1 v_{\infty}$.
The double-peaking discussed in Sect. 3.1 is strong, as well as a profile asymmetry
resulting from disk-occultation.
This latter effect, observed in O supergiants, would permit the constraint of the
wind acceleration just above the hydrostatic photosphere.
}
\end{figure}

Following NLTE model atmosphere results (Hillier 1988),
as well as the convention from the previous {\it lpv} analysis by LM,
let us now specify the extent of the LER to follow
a Gaussian form
\beq f(r) = \exp(-[(v(r)-v_e)/\Delta v_{e}]^2  ) \, ,
\eeq
where $\Delta v_{e}$ is the velocity range for emission about the velocity centroid $v_{e}$.
For an outer wind velocity centroid $v_{e} = 0.9 v_{\infty}$ (and a
relatively  narrow velocity range of $\Delta v_{e} = 0.1 v_{\infty}$),
Fig. 4 illustrates the resulting emission profiles for optically thin
(dashed) and optically thick (solid) lines, comparing again a fast
($\beta =1$; upper panel) and a slow ($\beta=4$; lower panel) acceleration.
The profiles are again all quite insensitive to the flow acceleration.
Note that the Gaussian form of the LER now imparts a rounding to the
edges of the flat portion for the thin lines, but for thick
lines the already curved form of the profile makes this additional
rounding less noticeable.
The vertical dotted lines mark the velocity centroid.
Note that in both cases this lies {\em outside} the width at
half-maximum, which is commonly used in observational analyses as a
rough measure of the LER velocity centroid;
for thick lines, such use of half-maximum leads to a substantial
underestimate of the characteristic velocity of the LER.
This can have important implications for the interpretation of {\it lpv}, as we
discuss below (Sect. 4).

Figure 5 illustrates corresponding profiles for a line formed in the
inner wind, with  $v_{e} = 0.3 v_{\infty}$.
For the optically thick lines,
the double peaked profiles are the result of the concave-up form of the
subshell emission, as discussed in Sect. 3.1;
the marked Blue/Red asymmetry seen in the $\beta=1$ case arises from
occultation of the redshifted emission from the hemisphere
behind the star.
Occultation also has a smaller, but still noticeable effect on the
red edge of the optically thin line, giving the overall profile a
modest blueward shift.
By contrast, occultation effects are much reduced in the $\beta=4$
models, simply because the corresponding radius for the fixed velocity
centroid $v_{e} = 0.3 v_{\infty}$ is now much further away from the star.

Such B/R asymmetry is commonly observed in O supergiants, and so
provides an useful constraint on the acceleration and velocity law
(Puls et al. 1996; Crowther et al. 2002).

\subsection{Effect of emission profile broadening}

To account for the effects of finite broadening (by thermal or
micro-turbulent motions), we next apply the ray-integration method of
Castor (1970; based on the Sobolev approximation (Sobolev 1960)),
with the narrow (delta function) emission now replaced
by a Gaussian of velocity width
$v_{\rm turb}$,
\beq
\phi (x) \sim \exp
\left [ -(x v_{\infty}/v_{\rm turb})^{2} \right ]
\, .
\eeq

\begin{figure}
\vspace{8cm}
\includegraphics{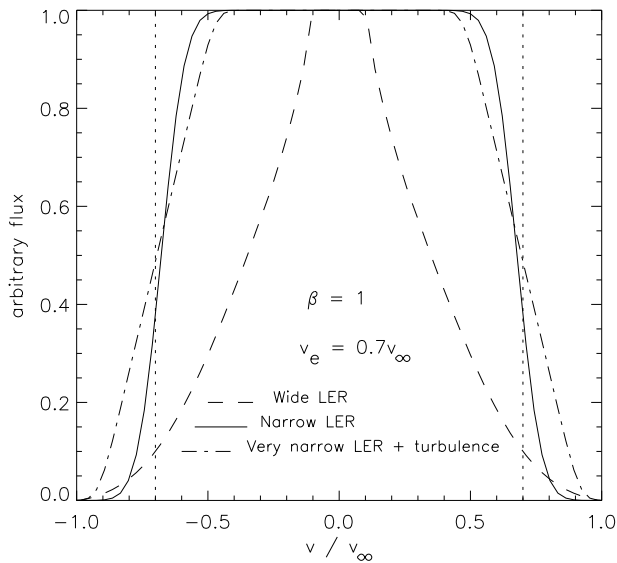}
\caption{
Optically thin emission profiles for models with $\beta=1$ and
velocity centroid $v_{e} = 0.7 v_{\infty}$, and three combinations of
LER velocity width $\Delta v_{e}$ and turbulent velocity $v_{\rm turb}$.
For cases without turbulence ($v_{\rm turb}=0$), the solid and dashed
curves show results for a wide ($\Delta v_{e} = 0.3 v_{\infty}$) and
narrow ($\Delta v_{e} = 0.1 v_{\infty}$) LER.
The dot-dashed curve is for a very narrow LER
($\Delta v_{e} = 0.05 v_{\infty}$) but with substantial turbulence,
$v_{\rm turb} = 0.2 v_{\infty}$.
The vertical dotted line marks the velocity centroid $v_{e}$.
See Sect. 3.3 for further discussion.
}
\end{figure}

Figure 6 shows the optically thin emission profiles for models with a common
velocity centroid ($v_{e} = 0.7 v_{\infty}$), but with three distinct
combinations of LER width and turbulent velocity.
The solid and dashed curves compare a narrow ($\Delta v_{e} = 0.1
v_{\infty}$) and wide ($\Delta v_{e} = 0.3 v_{\infty}$) LER,
without any turbulence ($v_{\rm turb}=0$).
Note that the latter gives a centrally peaked profile with very
extended line wings;
again the half-maximum width thus gives a severe underestimate of the
LER velocity centroid.
The dot-dashed curve shows the profile of a very narrow LER ($\Delta
v_{e} = 0.05 v_{\infty}$), but now with a substantial turbulent
velocity ($v_{\rm turb} =0.2 v_{\infty}$);
this turbulence further smoothes the edge of the profile, but
{\em does not substantially increase the overall profile width},
even though such a turbulent velocity is at the upper limit of what
is inferred in observational analyses (Robert 1992; LM).
This latter point plays an important role in our revision of the LM
analysis of {\it lpv} (see Sect. 4).

We can conclude from this section that a detailed numerical approach is required
for the determination of the LER characteristics of broad and rounded profiles, since
these profiles may be subject to significant optical depth effects.
Observed WR emission profiles showing a central peaking and broad wings reflect
a line formation over an extended LER.
Because they do not show any double-peaking, these lines are most likely
optically-thin.
The velocity at half-maximum is a poor indicator of the LER velocity centroid,
due to the bias of profile flux in favor of inner wind regions.
Broad flat top profiles are an unambiguous sign of optically thin line formation
close to terminal velocity (otherwise they would be narrow with minute flat top portion)
and over restricted ranges in velocity (otherwise they would be peaked).
These latter signatures
are the preferred diagnostics for {\it lpv} studies.

\begin{figure*}
\vspace{14cm}
\includegraphics{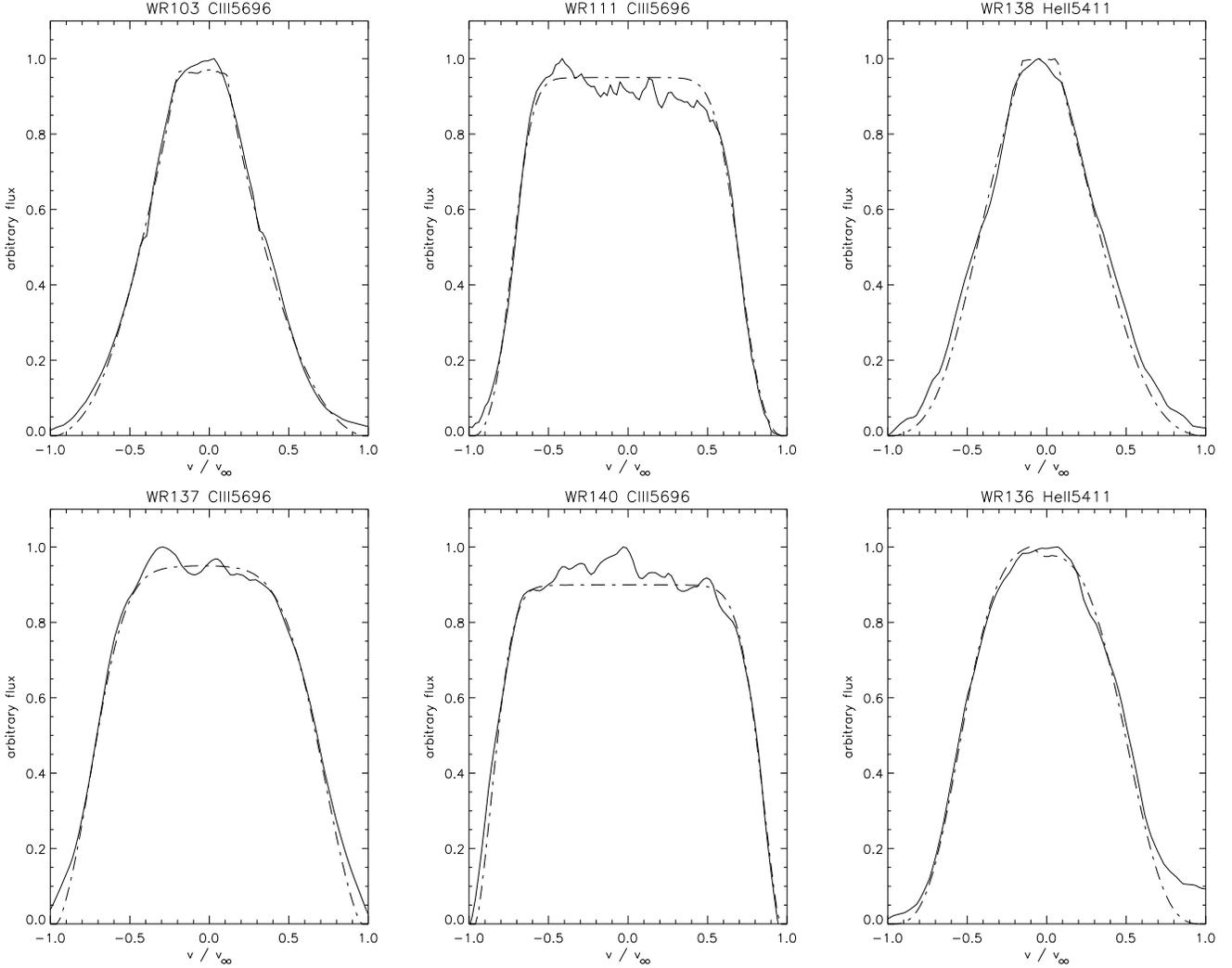}
\caption{
Observed profiles (solid curves) and best
fits (dot-dashed curves) in a chosen emission line
(He{\sc ii}\,5411\AA\,  or C{\sc iii}\,5696\AA) for six
WR stars discussed in LM. The intensity has been rescaled so that the
continuum is at zero and the maximum intensity is at unity. The wavelength
is in velocity units, scaled to the terminal velocity of the corresponding line,
as given in Table 1.
}
\end{figure*}

\subsection{Application to line profiles observed at a time snapshot}

\begin{table}
\caption[]{
Parameters used in profile fitting. The modeled line is
He{\sc ii}5411\AA\, for WR\,136 and WR\,138, and C{\sc iii}5696\AA\,
for all other cases.
To substantiate the terminal velocities used in this study (first column),
we compare values from LM (taken originally
from from Prinja, Barlow \& Howarth (1990),
labeled a), and from various
other studies (b: Eennens \& Williams 1994; c:
Hillier \& Miller 1998; d: Crowther \& Smith 1996; e: Howarth \&
Schmutz 1992).
Note that all values typically agree within ca. 100 \kms.
The last column denotes whether the fit requires an optically thick or
thin line.
}
\begin{center}
\begin{tabular}{l|ccc|ccc}
\hline
WR  & \multicolumn{3}{c}{$v_{\infty}$}&  $v_{e}$   & $\Delta v_e$ &$\tau_{line}$ \\
\hline
$\#$ &  \multicolumn{3}{c}{\kms} & $v_{\infty}$ &  $v_{\infty}$ &      \\
\hline
    & This work & LM$^a$ & diverse &  & &  \\
\hline
103 &  1200& 1190 & 1100$^b$  & .7 & .3 & thin \\
111 &  2400& 2415 & 2300$^c$  & .77 &.14 &  thin \\
135 &  1525& 1405 & 1525$^b$  &  .75&.2 & thin \\
136 &  1850& 1605 & 1750$^d$  &  .9&.27   & thick \\
137 &  1900& 1885 & 1900$^b$  &  .84&.22  & thin \\
138 &  1400& 1345 & 1400$^e$  & .65 &.25  & thin\\
140 &  2900& 2900 & 2870$^b$  &  .9& .15  & thin \\
\hline
\end{tabular}
\end{center}
\end{table}

We are now in a position to use the model described above
to assess the LER characteristics of time-snapshot observations of the 
recombination lines He{\sc ii}5411\AA\,
(WNs) and C{\sc iii}5696\AA\,  (WCs) for seven WR stars: WR\,103, WR\,111, WR\,135,
WR\,136, WR\,137, WR\,138 and WR\,140.
This sample was chosen to match that used by Robert (1992) for her
study of {\it lpv} in WR stars, but the short-time exposures (ca. 30 min.) we
use here were obtained independently by P. Crowther (p.c.).
(We exclude WR\,40 from our sample because its He{\sc ii} line shows a
narrow P-Cygni profile for which our method does not apply.)
Let us recapitulate the model assumptions.
We adopt a spherically-symmetric monotonically-expanding outflow.
Line emission occurs via recombination, thus possesses a density-square
dependence (see Eqn. 15).
No continuum opacity is included but allowance is made for variable
line-opacity.
Turbulence is treated as in Eqn. (20).
In Hillier (1988), a detailed description of the ionization stratification is
made for WR stars in general, showing that it is indeed realistic to use such a height
segregation for recombination lines in WR atmospheres (see also Dessart et al. 2000).
There is no explicit dependence of the emissivity on temperature, but it
implicitly enters our treatment via this height segregation of
wind ionization (Eqn. 19).
We show our fits to single-exposure line profiles for these objects in
Fig. 7, which shows the observed profiles (solid curves), together with our
model fits (dot-dashed curves).
(For clarity, we omit WR\,135 from the figure; its profile resembles that of WR\,111).
The model parameters are presented in Table 1.
The best fits are obtained by ``eye-ball'' estimate after iteration from a first
guess, modifying the various parameters: line opacity, centroid and extent
of the LER, without need to include any turbulence.

Note first that only for one case -- WR\,136 -- is it necessary to
invoke line optical depth effects;
this contrasts with the claim that such effects can be responsible for outstanding
fit discrepancies (LM).
For candidates other than WR\,136, we can
identify some lines with extended line wings
-- for which we infer a broad LER (WR\,103 and WR\,138) --
and others with near vertical line wings
-- for which we infer a narrow LER (the rest of the sample of stars);
overall, the velocity width spans the range
$\Delta v_e = 0.1-0.3 v_{\infty}$.

But a particularly important result is that the velocity centroids of the
LERs are all substantially higher than inferred by LM, in the range
$v_{e} = 0.65-0.9 v_{\infty}$.
Indeed, when these centroids are coupled with the inferred widths, i.e. $v_{e}+\Delta v_{e}$,
we see that the line formation region in all case extends essentially
all the way to the terminal speed, $v_{\infty}$.
This fact is manifest by the close correspondence between the maximum
base width of these optical emission lines and the terminal speeds inferred
by UV resonance lines.

Note that our detailed modeling is capable of fitting line profiles with a very
high level of accuracy since minor changes to model parameters lead to very different synthetic
line profiles; the fit quality is markedly degraded if either $v_{e}$
or $\Delta v_e$ are changed by even a few percents.
The central part of the synthetic profiles for WR\,103 and WR\,138 show minute variations
due to numerical inaccuracies, resulting from the strong emissivity gradient at the base of
the LER, while that of WR\,136 shows a slight Blue/Red asymmetry due to disk occultation
effects (Sect. 3.2 and Fig. 5);
In the present situations, such low-level irregularities, either real or from numerical error,
would disappear with even a modest turbulence.

With such information on the velocity in the formation region of emission lines with
observed {\it lpv},
we can now move on to using the wind acceleration derived from the migration of profile sub-peaks
to infer the overall scale of the wind velocity law, as characterized
by the value of $\beta \Rstar$.

\section{Application to inferred acceleration}

\begin{figure*}[!htp]
\vspace{12cm}
\includegraphics{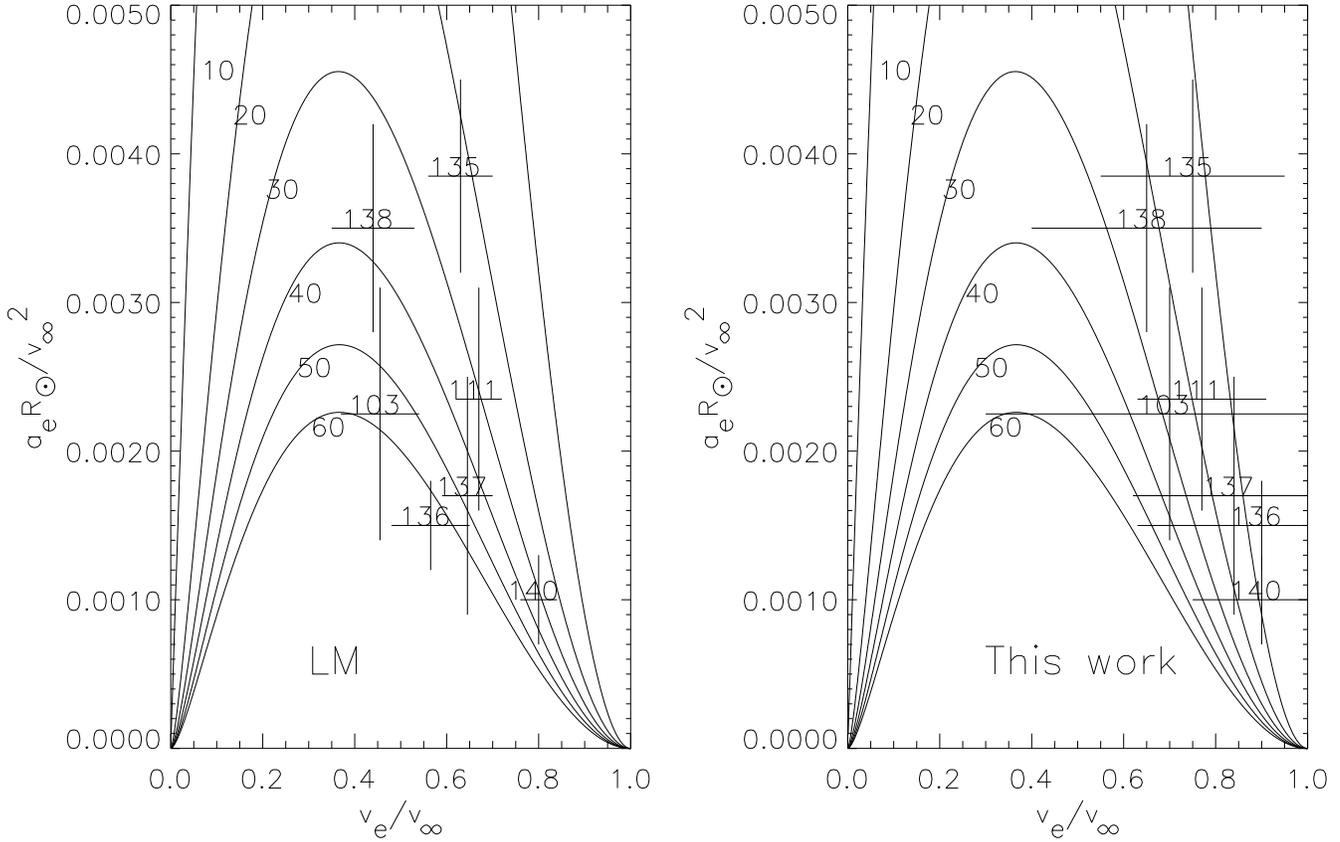}
\caption{
{\bf (Left)} A reproduction of Fig.~14 of LM. This figure depicts
the local wind acceleration a$_e$ (determined from a degradation function analysis)
at a given velocity v$_e$ in the stellar atmosphere for seven stars analyzed in LM (crosses).
The vertical error bar reflects an inaccuracy in the determination of the
acceleration from the DWEE model of LM, while the horizontal error bar reflects
the extent of, i.e. not an uncertainty in the position of, the LER in
velocity space. The over-plotted curves are the theoretical
a$_e$ versus v$_e$ for different values of $\beta R_{\ast}$.
{\bf (Right)} Id. but with the LER characteristics determined in this work, i.e.
crosses are shifted rightward by $.05-.25 v_{\infty}$ compared to LM, with
considerably broader velocity ranges of the emission sites.
}
\end{figure*}

For these program stars, LM conducted an extensive analysis of the {\it lpv}
observed by Robert (1992), inferring wind acceleration from the migration of
profile sub-peaks, based on a phenomenological Discrete Wind Emission
Element (DWEE) model.
Using the inversion technique of Brown et al. (1997), LM
estimated the LERs to extend over a narrow velocity range
($\Delta v_{e} \approx 0.05 v_{\infty}$)
located relatively deep in the wind
($v_e=0.4 - 0.8 v_{\infty}$),
with turbulence claimed to be the main profile broadening mechanism.

Such narrow LERs warranted taking a constant wind acceleration over
its limited range, a key assumption for the degradation function method used
to infer wind acceleration magnitudes.
Combining these with the inferred LER
velocities, LM concluded that the associated flow acceleration must extend over
very large length scales, $\beta \Rstar \approx 20 - 80~ \Rsun$.
These are much larger than the estimated hydrostatic core radii of WR stars
($< 5 ~\Rsun$, Hillier \& Miller 1999; Dessart et al. 2000, Crowther et al. 2002),
a result that challenges the standard radiative driving paradigm for these WR winds (Sects. 1--2).

By contrast, the LER characteristics derived here differ markedly from
those inferred by LM, with systematically higher velocity centroids
and larger widths, generally extending the outer range of the LER
to nearly the terminal speed.
Adopting the same acceleration magnitudes derived by LM, this
now implies much more modest length scales for the wind
acceleration.
This  follows simply from the fact that line formation in regions close
to terminal speed can be expected to exhibit relatively low
acceleration magnitudes.

These central results are illustrated in Fig. 8, wherein
the left panel reproduces Fig. 14 of LM,
while the right panel reconfigures the LER location and width
according to the results given in our Table 1.
Comparing the left to the right panel, note that the acceleration
values are the same, but the LER velocity centroids (crosses) are shifted rightward
(by ca. $0.1-0.3 v_{\infty}$) and the velocity ranges (horizontal
bars) are substantially broadened (typically more than doubled).
As in LM, the overplotted curves in both panels show the associated
variation of acceleration vs. speed for models with various
acceleration scales $\beta \Rstar$, labeled in units of $\Rsun$.

Note in particular that our LER parameters in the right panel are now
consistent with substantially smaller acceleration lengths
($10- 20 \, \Rsun$) than implied by the LM values in the left
panel ($20 - 60 \, \Rsun$).
These are still quite large, but perhaps now within reach of what
radiative driving can achieve if one accounts for multi-line
scattering (Springmann 1994; Gayley, Owocki, \& Cranmer 1995).

The origin of this difference lies solely in the new parameters derived for the
LER velocity location and width.
LM cited turbulence as the dominant source of broadening of WR line profiles;
but it seems that with LM's adopted narrow and deep LER,  line profiles with
extended wings can only be fitted by invoking very large turbulent velocities,
indeed much higher than those determined from their own {\it lpv} analysis.

Figure 9 illustrates these points for the sample case of the He{\sc ii}5411\AA\, 
line from WR~138, comparing the observed profile (solid curves)
with various models (dot-dashed curves).
The top panels show our models with no turbulence (left) and with
the {\it lpv}-derived turbulence (right);
both fits are quite good, though the turbulence introduces a smoothing
and a moderately improved agreement.
The bottom panels show LM models with this same {\it lpv} turbulence (left)
and with the much larger turbulence needed to fit the observed
profile (right);
the former is the turbulence value cited by LM, but we clearly see
that this does not provide an acceptable profile fit when combined
with their inferred LER characteristics.
Instead, a quite unrealistic turbulence is needed to fit the line.

\begin{figure}[!htp]
\vspace{9cm}
\includegraphics{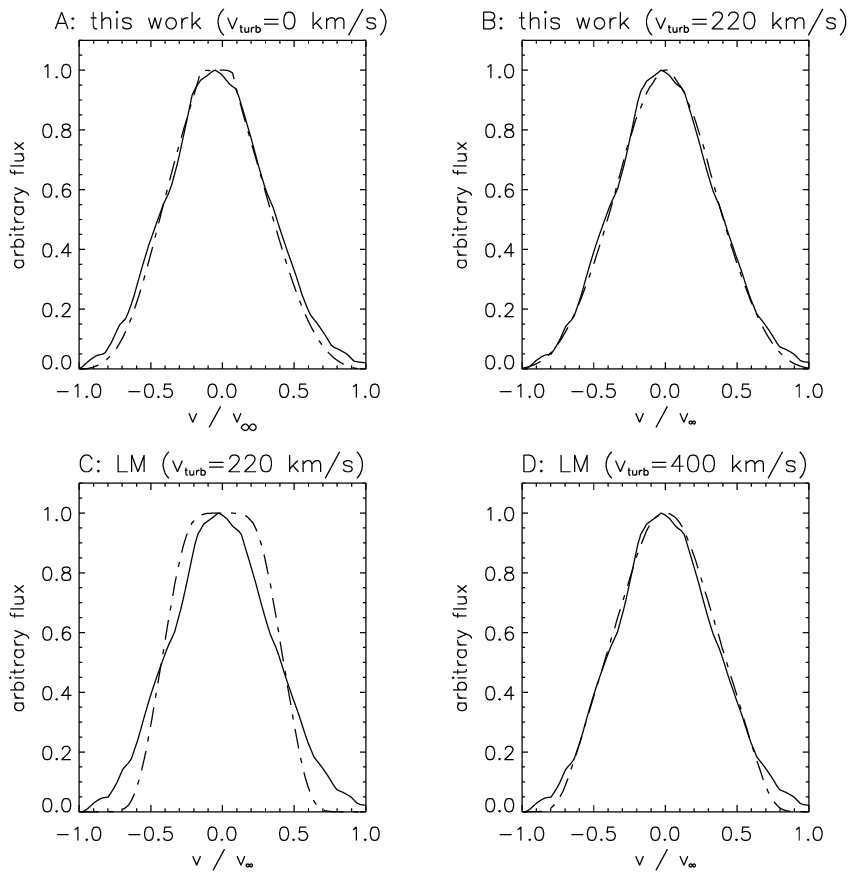}
\caption{
{\bf Top:}
Observed He{\sc ii}5411\AA\,  profile of WR\,138 (solid curves), 
compared with models using the parameters of the present study 
(dot-dashed curves), without (A) and with (B, $v_{\rm turb}=220$ \kms) 
turbulence.
{\bf Bottom:} Same but with the parameters of LM (C), and with a
turbulent velocity twice the value determined from the {\it lpv}
analysis (D, $v_{\rm turb}=400$ \kms).
}
\end{figure}

In the present models with the line emission formed over a {\em range}
of velocities, the emission profile is rounded even without
turbulence (Fig. 9, panel A), and so the addition of a modest
turbulence leads to only a modest additional broadening without much
affecting the overall emission profile shape (cf. panels A and B of Fig. 9).
By contrast, in the LM models, in which the LER is assumed to be at
nearly a single velocity, the intrinsic profile has a more flat-topped,
box shape, even with the modest level of turbulence (i.e. $v_{\rm turb} =
220$ \kms; Fig. 9, panel C).
Because of the relatively steep gradient in emission in going from the
core to wing, the stronger level of turbulence ($v_{\rm turb} = 400$ \kms
now has the general  effect of {\it redistributing} flux from near the edges
of the flat-top into the line wings; the overall effect then is to broaden
the base (wing) of the profile, but round-off and narrow the emission core.
An essential point is that the narrow range of velocity assumed by LM
for the LER requires just such a large and physically dubious level
of turbulence to produce the observed width and form of the line
profile, whereas this can be produced naturally without invoking
turbulence if one allows for the LER to span a modest range in outward
flow speed.

This exercise supports the view (Hillier 1988)
that WR line broadening results primarily from the finite extent of the LER in
velocity space, and not from turbulent broadening.
These changes relative to the LER properties used by LM may not have
much effect on the acceleration {\em magnitudes} derived in  their {\it lpv} analysis (see Sect. 5),
but they do imply a substantial downward revision (by a factor of two
or more) in their quoted acceleration {\em scales},
i.e. from  20-80~$\Rsun$ to 10-20~$\Rsun$.

\section{Wind acceleration inferred from synthetic {\it \bf lpv}}

  The migrating sub-peaks observed in {\it lpv} are understood to stem from the variable
emissivity of the structured wind accelerating through the LER (Sect. 1).
Lucy (1984) and Owocki \& Rybicki (1984) proposed that this wind structure
could result from the strong instability to small-scale perturbations of
the line-driving mechanism.
The non-linear evolution of a line-driven wind subject
to such an instability was studied for the first time in a
1D radiation-hydrodynamics computation by Owocki, Castor, \& Rybicki (1988),
showing that the outflow relaxes into a highly-developed non-monotonic radial
velocity field, whose associated shocks lead to radially-confined slow high
density compressions separated by extended fast rarefied regions.
We present these properties for one such simulation of an O-star wind (with suitable
parameters for the supergiant star $\zeta$ Puppis) in the top panels of
Fig. 10 for the radial velocity (left) and density (right)
as a function of radius.
When plotted in a Lagrangian coordinate (bottom panels, Dessart \& Owocki 2002b),
the bulk of the material follows a velocity, which, when
time-averaged, matches to within a few percents the CAK velocity law with
a $\beta$-exponent value of 0.8.
In other words, only a negligible fraction of the total wind mass departs from
such a CAK velocity, contributing relatively little flux to (density-squared dependent)
emission profiles.
Thus, this property justifies the use of a standard $\beta$-velocity law
for the wind structures, as done in LM for the modeling of {\it lpv}.

   However, because the lines used by LM have a finite rather than a narrow LER (Sect. 4),
assuming a fixed wind acceleration may lead to systematic errors in the derived
acceleration magnitudes.
Indeed, in the case, e.g., of a wide LER, most of the profile flux
originates from the low-velocity high-density regions (Sect. 3), which for a fully
clumped wind, tends to favor the variability signal of those deep wind regions.
   In this section, to circumvent such a kind of biases, we present a more direct
comparison between observed and synthetic {\it lpv}.
To do so, we compute the optically-thin line emissivity arising from a 3D wind assumed to be
composed of independent star-centered cones characterized by a fixed angular extent.
The crux of this so-called ''Patch'' method (described in great details in Dessart \& Owocki (2002a),
hereafter DO) is to adopt, for each cone, a
purely 1D wind layout as computed by radiation hydrodynamics simulations of the radiative instability.
A representative snapshot of the height variation of the velocity and density is shown in Fig. 10,
both in Eulerian and Lagrangian coordinates.
Thus, while the method retains some 1D features as far as the gas is concerned, it allows us
to introduce a complete wind anisotropy, which, as  illustrated in DO, is required by the lack of
emission profile variations in the line center region.
Moreover, Dessart \& Owocki (2002b) interpreted the velocity-scale of variable profile sub-peaks
at line-center as arising from the lateral geometrical broadening of wind structures.
They showed that with a patch-size of 3 $\deg$, one obtains a satisfactory fit to the observed
sub-peak velocity width at line center as well as the fluctuation magnitude of observed {\it lpv}.
Thus, the same value of 3$\deg$ is used here.

\begin{figure*}[!htp]
\vspace{12cm}
\includegraphics{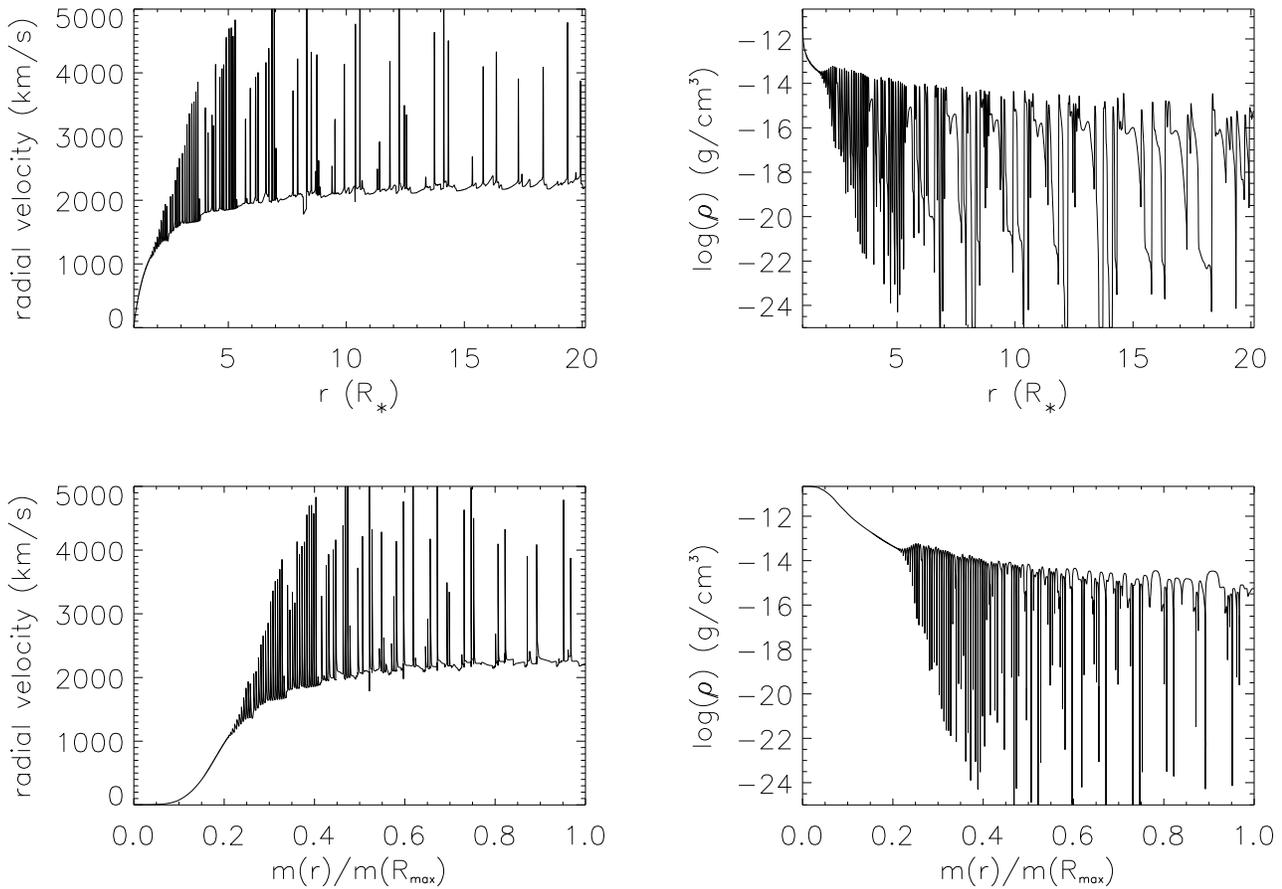}
\caption{
Single time-snapshot from the hydrodynamics simulation of the radiative
instability, showing the velocity (left column) and log-density (right column),
in Eulerian (top) and Lagrangian (bottom) coordinates.
}
\end{figure*}

In the following two sections, we use the Patch method together with
the hydrodynamical simulations of the radiative instability presented in Fig. 10 to
compute synthetic {\it lpv} of WR and O stars.
Strictly speaking, such hydrodynamical simulations only apply to winds driven {\it via}
single line-scattering, i.e. O star winds, and not to WR winds whose larger mass loss rates
require the treatment of multiple line-scattering (Gayley, Owocki, \& Cranmer 1995).
Nonetheless, both are understood to be line-driven and believed to share similar dynamical
properties, differences arising from the height location of line-emission:
all O-star optical diagnostics originate close to the hydrostatic core radius
(below ca. 2-3 R$_{\ast}$, Hillier et al. 2002) while those of WRs subsist out to large radii.
It is {\it via} such a height segregation that we now study such a synthetic {\it lpv} for
WR and O star emission lines.

\subsection{Measured sub-peak acceleration from synthetic {\it lpv}: WR-star case}

\begin{figure*}[!htp]
\vspace{19cm}
\includegraphics{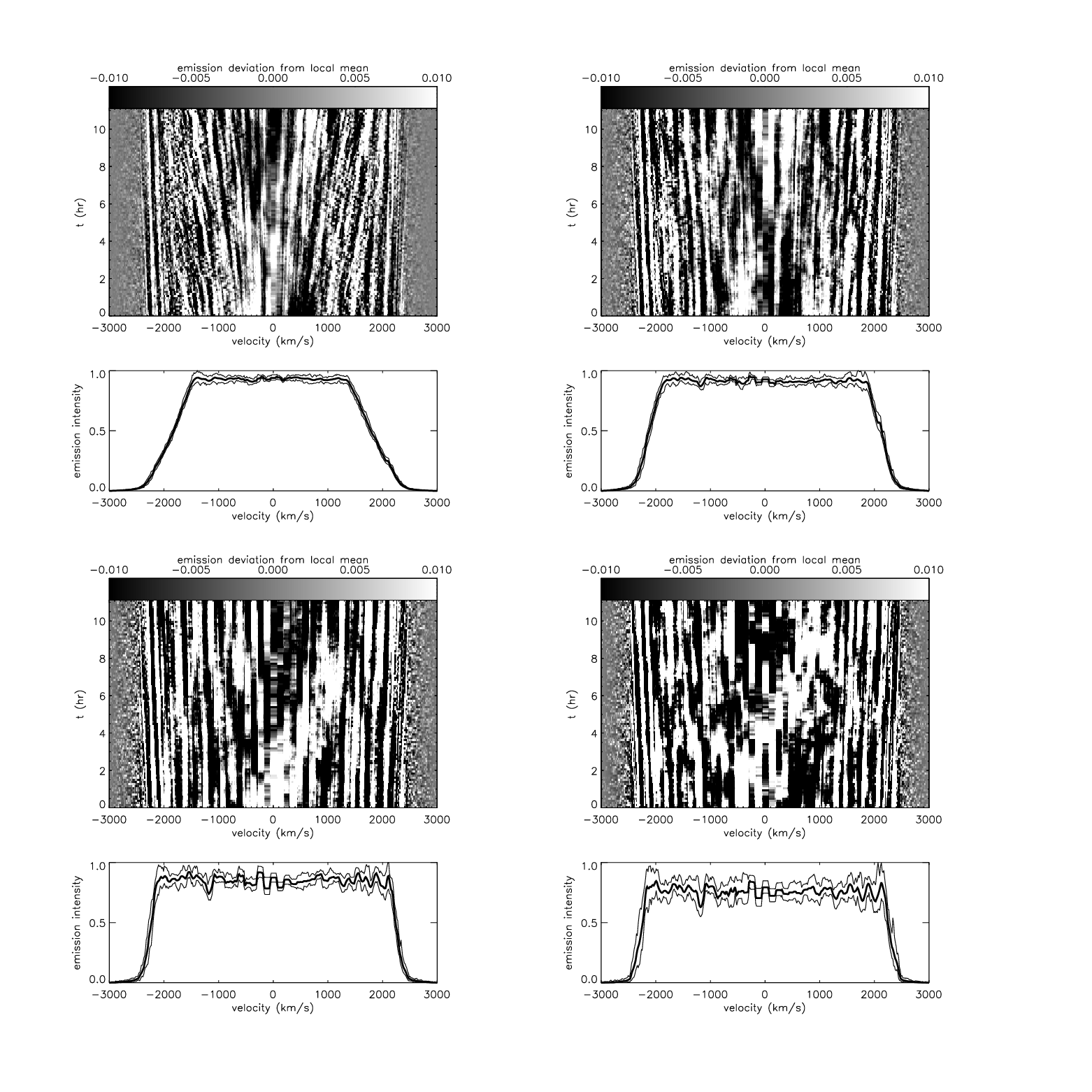}
\caption{
Montage of synthetic {\it lpv} based on radiation-hydrodynamics simulations of the
line-driving instability for four optically thin emission lines whose
LER extend out to 20$R_{\ast}$ but start at different depths: 2.5 and 5.0 $R_{\ast}$
(top left and right panels) and 10.0 and 15.0 $R_{\ast}$ (bottom left and right
panels).
To enhance the visibility of migrating sub-peaks in the grayscales, the local maximum
flux deviation is limited to 1\%.
}
\end{figure*}

We stressed in the previous sections the importance of using a reliable
LER characterization for the interpretation of {\it lpv}.
In particular, we found that in general, the lines studied by LM formed
over a finite LER that extended out to terminal velocity.
Hence, here, for all profile computations, we set the outer LER radius at
the maximum radius of the hydrodynamical simulation of $20 R_{\ast}$,
corresponding to the terminal velocity of the flow within a few percents.
We then vary the inner radius of the LER to reproduce the range of observed
widths of {\it lpv} diagnostics, covering the values $r_{l} = 2.5, 5, 10$ and
$15\,R_{\ast}$.
We show a sequence of synthetic grayscale time-series in Fig. 11, with increasing
$r_{l}$ value from top to bottom, and left to right.

In the top-left panel, we see the presence of sub-peaks with a range
of acceleration magnitudes. Around line of sight velocities of
1000 \kms and -1000 \kms,
we can estimate values of ca. 50 \kms/hr,
which due to a line of sight angle cosine weight of about a half,
correspond to an higher intrinsic wind acceleration of ca. 30 m\,s$^{-2}$.
In the line wing, the sub-peak migration magnitude is systematically lower, ca.
20 \kms/hr and since these profile patterns arise from wind emitting
structures advecting directly towards us, the sub-peak acceleration corresponds
directly to the intrinsic wind acceleration, i.e. of ca. 8 m\,s$^{-2}$.
This is consistent with the fact that the LER is wide, covering a large range of
intrinsic wind accelerations. Despite additional geometrical effects, this
leads to a decreasing sub-peak migration magnitude from line center to line wing.

\begin{figure*}[!htp]
\vspace{10cm}
\includegraphics{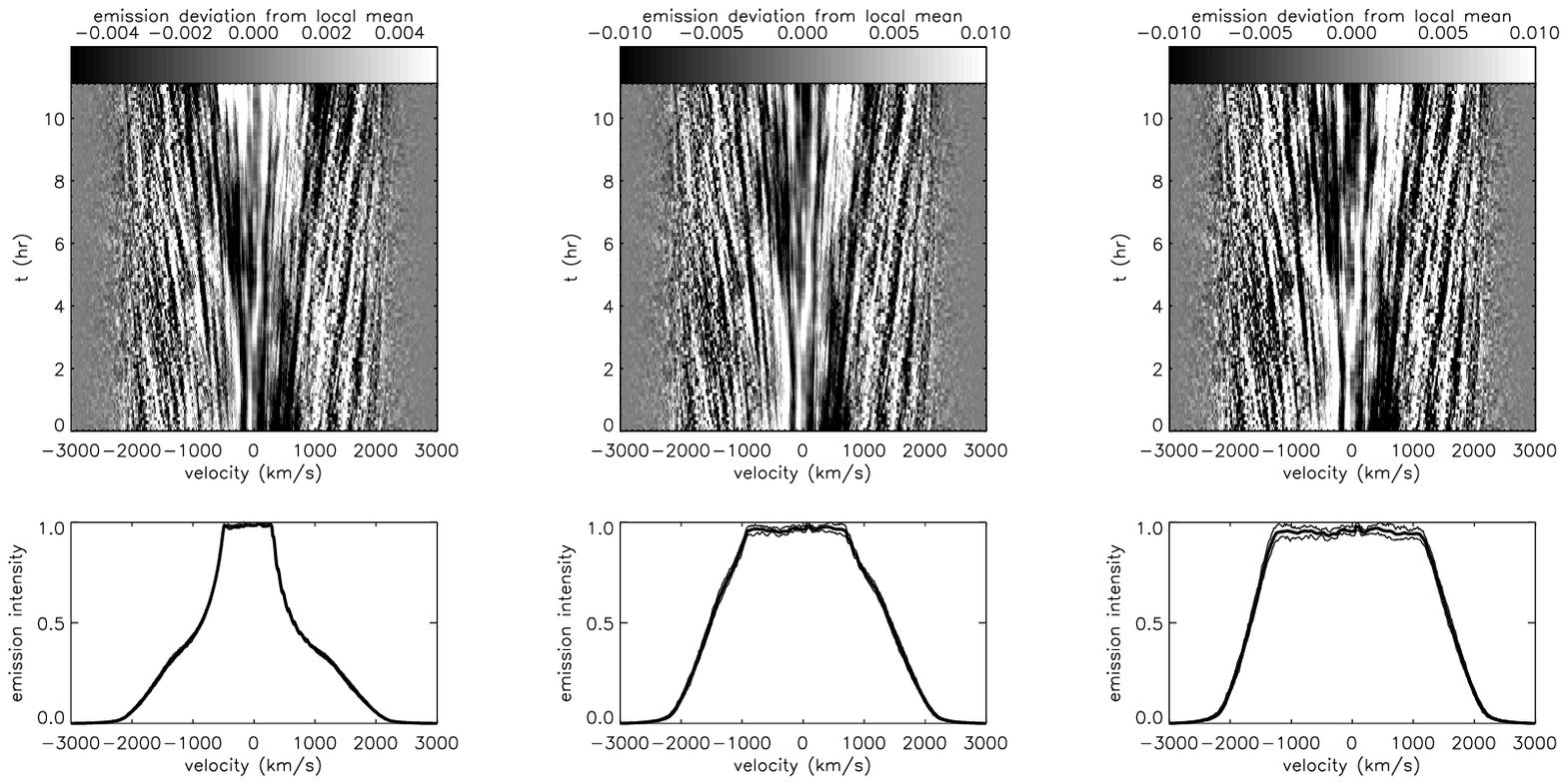}
\caption{
Same as previous figure but for 3 emission line profiles
whose LER starts at 1.2, 1.5 and 2.0 $R_{\ast}$ (left to right)
but extends out to the same outer radius, i.e. 10.0 $R_{\ast}$.
}
\end{figure*}

As we move the inner edge of the LER to larger radii (subsequent three panels),
the fast migration of sub-peaks in the flat part of the profile is at first
less conspicuous (top-right panel), before disappearing altogether for outer
wind forming lines (bottom panels).
In these latter cases, the LER covers a narrow range of velocities and is
well characterized by a unique wind
acceleration directly measurable from line-wing subpeaks. Contrary to the case
of a deep forming line, the profile migration now increases from line center
to line wing. Its magnitude is small, consistent with the fact that the LER
now probes wind regions that are within 10\% of the terminal velocity.

Interestingly, the simulation  corresponding to a line whose LER starts at
2.5 R$_{\ast}$ (top-left panel) reproduces well both the temporal
evolution as well as the time-averaged line emission observed for CIII5696\AA\,
for WR\,111.
In particular, the migration magnitude of ca. 30 m\,s$^{-2}$ is
in good agreement with values given in LM as well as the observed sub-peak
migration in the flat part of the profile.
Note that the migrating subpeaks
appearing in the line wing are unobserved, a likely consequence of the fact
that 1D radiation hydrodynamics simulations underestimate the magnitude
of radial velocity dispersion.
But the more dramatic drawback from our
model is that we use a stellar radius of 19\,R$_{\ast}$, while the
hydrostatic core radius of WR\,111 is expected to be a factor of 2--10 smaller
(Hillier \& Miller 1999).
Hence, with such small core radii, the inferred WR wind acceleration length
scale is still significantly more extended than expected from radiation
driven wind theory which at present favors a ``standard'' velocity law
with $\beta$ of ca. 1--2 at most.

\subsection{Measured sub-peak acceleration from synthetic {\it lpv}: O-star case}

  In the previous section, we chose an inner radius for the
LER compatible with the generally large emission line profile
widths of WR stars.
By contrast, in O-stars, emission line formation starts right
at the base of the outflow, resulting in much narrower emission
profiles, usually quite pointed rather than flat-topped.
Thus because the whole wind is optically thin, a given emission
line carries information on all wind heights, allowing us to
constrain the inner wind acceleration from single-exposure line
profiles (Sect. 3.2; Puls et al. 1996; Crowther et al. 2002), as
well as from {\it lpv} (Eversberg, L\'epine, \& Moffat 1998).

  To model the {\it lpv} of O-star winds, we use the same approach
as in Sect. 5.1, this time setting the inner edge of the LER to
very small wind heights, covering the values 1.2, 1.5 and
2.0 R$_{\ast}$ (left to right panels in Fig. 11).
The migration magnitude of profile sub-peaks is quite comparable
between all three simulations, and only slightly higher that for
the situation where the LER starts at 2.5 R$_{\ast}$ (Sect. 5.1).
In our model (Fig. 10), although the wind acceleration magnitude increases towards the
wind base, the wind ceases to be clumped below ca. 2 R$_{\ast}$.
Hence, no profile variations arise from wind emission at these
depths, invalidating the use of our method based on {\it lpv} to infer
the wind acceleration.

  But, although the {\it lpv} pattern is very similar in these three
cases, the time-averaged line profile changes dramatically as
the LER inner radius decreases.
Indeed, the line-wing shape presents a very clear kink, at a
profile location which corresponds to the wind velocity at which
clumping appears.
The emissivity of the {\it lpv} diagnostics being density-squared
dependent, it is suddenly enhanced when going from the unclumped
inner wind regions through the clumped outer wind ones.
Thus, if O-star wind clumping only started at a sizeable fraction
of the terminal velocity, this feature would be observable.

  However, to our knowledge, such a profile kink has not been seen in
O-star spectra which might be supporting the idea that their winds are clumped
even in the inner wind region.
The observed {\it lpv} for the O supergiant star $\zeta$ Puppis
(Eversberg, L\'epine \& Moffat 1998) also suggested that the pattern of
migrating subpeaks was compatible with a wind clumping starting right
at the wind base.
Moreover, Prinja et al. (2001) reported a B/R asymmetry in the {\it lpv}
of HD\,152408 (O8:Iafpe or WN9ha), with the migration magnitude of
profile sub-peaks twice as small in the red part than in the
blue part of the line profile.
In Sect. 3.2, we showed that disk occultation can lead to a significant
flux deficit in the red part of the profile. But if this deficit is
combined to a B/R dichotomy in {\it lpv}, it also implies that the occulted
innermost wind regions are also clumped.
Thus, these various elements support the spectroscopic analysis of
O supergiant stars (Hillier et al. 2002, 2003; Crowther et al. 2002) which
require that the wind clumping starts right from the base.

\section{Summary and future work}

\begin{figure*}[!htp]
\vspace{8cm}
\includegraphics{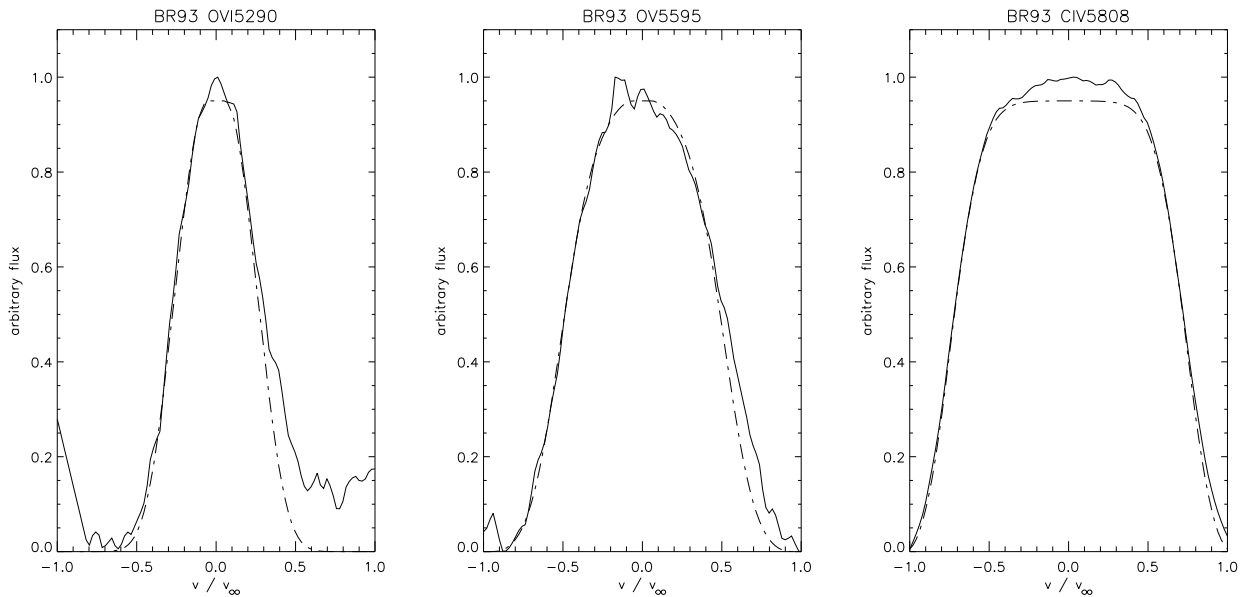}
\caption{
Optical emission lines of O{\sc vi}5290\AA, O{\sc v}5595\AA\,
and C{\sc iv}5808\AA\, observed in the Br93 spectrum. Superposed on each
profile is a satisfactory fit with the model described in Sect. 3.3.
The velocity centroid $v_e$ of the LER for the three modeled lines,
going from left to right, are 40\%, 65\% and 85\% of $v_{\infty}$, while the
velocity range of the LER is 20\% of $v_{\infty}$ for each.
No line opacity needed to be accounted for in order to reproduce the profile.
Some turbulence is included: v$_{\rm turb}$ $\sim$ 400 \kms, which corresponds
to 10\% of the terminal velocity.
This illustrates why WO stars represent ideal candidates to reveal 
{\it differential wind acceleration} from {\it lpv}.
}
\end{figure*}

We have reexamined the LER characteristics of WR {\it lpv} diagnostics (Robert 1992).
We find that, in general, these lines are optically thin, showing a profile
broadening that results from the finite extent of the LER rather than from turbulence.
Furthermore, we find that they form much closer to terminal velocity
than proposed in previous studies (LM), leading to a downward revision of the WR wind
acceleration length scales $\beta R_{\ast}$ from $20-80 \, R_{\odot}$ to
$10-20 \, R_{\odot}$.
Based on 1D radiation hydrodynamics simulations of the radiative
instability and the ``Patch'' method, we compute synthetic {\it lpv}
for C{\sc iii}5696\AA\  for WR\,111. We are able to reproduce
well the observed {\it lpv} for that star, although our model uses an
O-star core radius.
Both evolutionary and spectroscopic models predict small WR
core radii, a few times the solar value, which thus makes WR wind
acceleration length scales still considerably extended, although
not to the extent proposed by LM.
Forthcoming VLTI-AMBER observations will hopefully give an
accurate evaluation of the spatial scales of WR stars and their
associated winds, thereby setting the magnitude of the discrepancy
between the inferred WR wind acceleration from observations and that
expected from radiation driven wind theory.

In our synthetic {\it lpv} simulations for optically thin O-star winds,
we find a clear signature for the onset velocity of wind clumping,
which results from the strong sensitivity of recombination line
emissivity to variations in wind density.
This introduces a kink in the line-wing profile shape,
making the clumped outer wind regions relatively more emissive than
the smooth inner wind ones.
The absence of such a kink in even high-resolution spectroscopic
observations of O supergiants together with results from {\it lpv}
studies of such stars (Eversberg, L\'epine, \& Moffat 1998; Prinja
et al. 2001) suggest that indeed, wind clumping may well start right from
the hydrostatic base in O supergiant stars.

Finally, the WR-{\it lpv} dataset obtained so far uses diagnostic lines that
form close to terminal velocity; such lines contain only poor information
on the overall wind acceleration.
Using optically thinner winds would yield a more valuable information on the wind
acceleration.
This is the case for OB objects but their wind density is however so low that 
no line emission arises from above ca. 2-3 R$_{\ast}$ (Hillier et al. 2002), 
providing no information on the outer wind dynamics.
On the contrary, WO star winds have not only a low optical depth in the visual 
band but they also show strong optical emission lines of markedly different 
velocity widths, resulting from a strong wind ionization stratification 
(Crowther et al. 2000).
Indeed, using the model of Sect. 3, we find that optical emission lines of 
oxygen and carbon form in regions located between 40\% to 100\% of the 
terminal velocity (Fig. 13).
Thus, WO stars constitute excellent targets for {\it lpv} studies
to provide further observational constraints on the structure and dynamics 
of WR star winds.

\begin{acknowledgements}

SPO acknowledges support of NASA grant NAGW5-3530 and NSF grant AST-0097983, 
awarded to the University of Delaware.

\end{acknowledgements}

\end{document}